\documentclass[10pt,twocolumn]{article}

\usepackage{graphicx}
\usepackage{algorithm}
\usepackage{algpseudocode}
\usepackage{url}
\usepackage{verbatim}
\usepackage{listings}
\usepackage{color}
\usepackage{tabularx}
\usepackage{float}
\usepackage{caption}
\usepackage{subcaption}
\usepackage{mathptm}
\usepackage{booktabs}
\usepackage{multirow}

\title{Machine Learning Based Auto-tuning for \\
Enhanced OpenCL Performance Portability\footnote{This is a pre-print version an article to be published in the Proceedings of the 2015 IEEE International Parallel and Distributed Processing Symposium Workshops (IPDPSW). For personal use only.}}

\author{Thomas L. Falch and Anne C. Elster \\
Department of Computer and Information Science\\
Norwegian University of Science and Technology\\
Trondheim, Norway\\
Email: \{thomafal, elster\}@idi.ntnu.no
}

\date{}

\begin{document}

\lstset{basicstyle=\scriptsize\ttfamily,captionpos=b,language=C,morekeywords={complex}}

\maketitle

\begin{abstract}
    Heterogeneous computing, which combines devices with different architectures, is rising 
in popularity, and promises increased performance combined with reduced energy consumption.
OpenCL has been proposed as a standard for programing such systems, and  
offers functional portability. It does, however, suffer from poor performance portability, code tuned for 
one device must be re-tuned to achieve good performance on another device. In this paper, we use 
machine learning-based auto-tuning to address this problem. Benchmarks are run on a random subset 
of the entire tuning parameter configuration space, and the results are used to build an artificial neural 
network based model. The model can then be used to find interesting parts of the parameter space for 
further search. We evaluate our method with different benchmarks, on several devices, including an Intel i7 3770 CPU, 
an Nvidia K40 GPU and an AMD Radeon HD 7970 GPU. Our model achieves a mean relative error as low as 
6.1\%, and is able to find configurations as little as 1.3\% worse than the global minimum.
\end{abstract}


\section{Introduction}

One of the most popular heterogeneous platforms 
today is a latency optimized CPU with a few, high single thread performance cores, combined with 
one or more throughput optimized GPUs with many, slower cores, for high parallel performance.

While such systems offer high theoretical performance, programming them remains challenging. One 
notable issue is code portability. To improve the situation, OpenCL\cite{OPENCL} was 
proposed. Programs written in OpenCL can be executed on any 
device supporting the standard. Currently this includes CPUs and GPUs from AMD, Intel and 
Nvidia as well as devices from other vendors.

Although OpenCL offers functional portability, i.e. OpenCL code will run correctly on 
different devices, it does not offer performance portability. Instead, code must be re-tuned for each new device it is 
executed on. The problem of performance portability is not new or not tied to 
OpenCL. For instance, code tuned for one CPU will often require 
re-tuning if ported to a new generation of CPUs, or a CPU from a different vendor. However, this 
problem is exacerbated with OpenCL, since it supports a larger variety of devices, with 
more diverse architectures.

Auto-tuning may be used to overcome this issue. In its simplest form, 
auto-tuning involves automatically measuring the performance of several candidate 
implementations, and then picking the best one. Auto-tuning can be divided into two types:
empirical, and model driven. In empirical auto-tuning, all possible candidate implementations are 
evaluated in order to find the best one. While this guarantees that the optimal implementation can 
be found, it can be very slow if there is a large number of candidates. Model-driven auto-tuning 
attempts to solve this problem by introducing a performance model which is used to find a 
subset of promising candidates, which are then evaluated. While this reduces the time required 
for the auto-tuning, the results depend heavily on the quality of the performance model, 
which furthermore can be difficult and time costly to develop.

There is also a third approach. Instead of manually deriving an analytical performance model, it 
can be built automatically instead, using machine leaning methods. A random set of 
candidate implementations are executed, and the measured execution times are used to learn a 
statistical model. This model is then used to pick promising candidates for evaluation, as with 
traditional model driven auto-tuning.

In this paper, we show how to use machine learning based auto-tuning to re-tune OpenCL code to 
different devices. We achieve good performance without evaluating a large number of 
candidates, or manually build a performance model.

The remainder of this paper is structured as follows: The next section  
demonstrates the need for solutions to the problem of poor OpenCL performance 
portability. Section~\ref{related} provides an overview of related work, while 
Section~\ref{background} contains background information on heterogeneous computing, OpenCL and 
machine learning. Our method is described in Section~\ref{work}. Results are presented in 
Section~\ref{results}, and discussed in Section~\ref{discussion}. Finally, Section~\ref{conclusion} 
concludes, and outlines possible future work.

\section{Motivational Example}
\begin{figure}[htpb]
    \centering
    \includegraphics[width=0.45\textwidth]{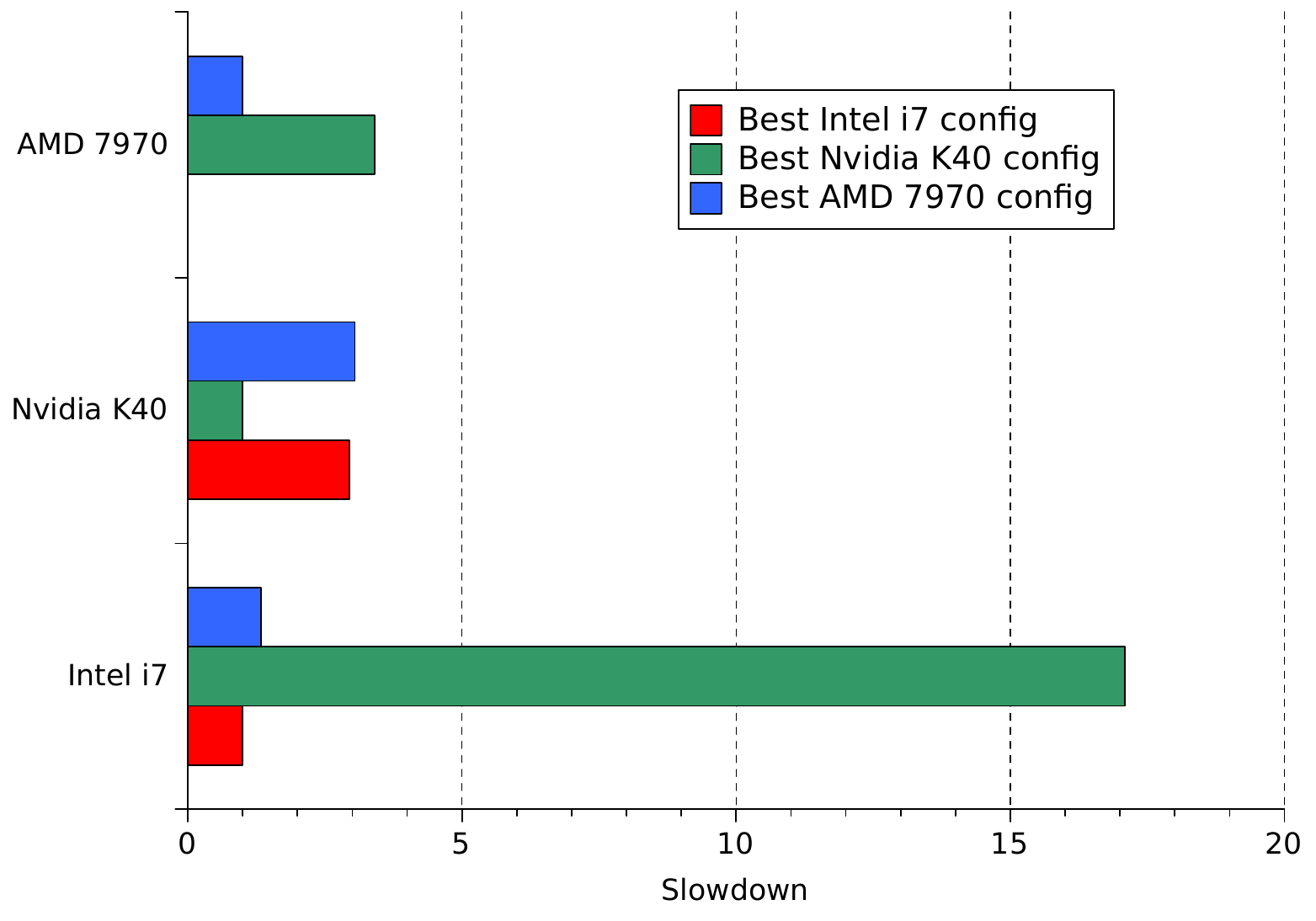}
    \caption{The slowdown of three different configurations compared to the best configuration for 
the different devices.}
\label{motivation}
\end{figure}

To illustrate the poor performance portability of OpenCL, we ran a convolution benchmark (described
in Table~\ref{benchmarks}) on three common devices, an Intel i7 3770 CPU, an Nvidia K40 GPU and an
AMD Radeon HD 7970 GPU. The benchmark has a number of tuning parameters, such as the work-group
size, and whether or not to apply various potential optimizations (described in
Table~\ref{param_table}). Since the architectures of the devices are different, we expect that the
best tuning parameter configurations will also be different. We confirmed
this by exhaustively trying all possible configurations for all three devices, thereby finding the
best Intel configuration, the best Nvidia configuration and the best AMD configuration, which all
differed from each other. We then measured the performance of these three parameter configurations
on all the devices. The results are shown in Figure~\ref{motivation}.

As we can see, using the "wrong" configuration can seriously degrade the performance, even when that configuration
is the best one for another device. For instance, using the best Nvidia configuration on the Intel i7 resulted in
a slowdown of 17.1 compared to the best Intel configuration. Even between the two GPUs, which have a
more similar architecture, the problem persisted. Using the best AMD configuration on the Nvidia
card, or the best Nvidia configuration on the AMD card both resulted in slowdowns of approximately
3.

This clearly demonstrates the need for re-tuning OpenCL code when executing it on a new 
device. Furthermore, resorting to exhaustive search, as we did here, is in general not practical or 
even possible since the configurations spaces can be very large.

\section{Related Work}
\label{related}

Auto-tuning is a well established technique, which has been successfully applied in a number of 
widely used high performance libraries, such as FFTW\cite{FFTW,MEYER1} for fast Fourier transforms, 
OSKI\cite{OSKI} for sparse matrices and ATLAS\cite{ATLAS} for linear algebra\cite{JENSEN}.

There are also been examples of application specific empirical auto tuning on GPUs, e.g. for stencil
computations \cite{ZHANG2}, matrix multiplication \cite{LI} and FFTs \cite{NUKADA}.
Furthermore, analytical performance models for GPUs and heterogeneous systems have been 
developed\cite{MEYER,MEYER2,MEYER3,HONG} and used for auto-tuning \cite{YOTOV}.

Much work has been done on machine learning based auto-tuning, e.g. to determine
loop unroll factors\cite{STEPHENSON}, whether to perform SIMD vectorization\cite{TROUVE} and general
compiler optimizations \cite{MILEPOST}. Kulkarni et al.\cite{KULKARNI} developed a method to determine
a good ordering of the compiler optimization phases, on a per function basis. Their method uses a
neural network to determine the best optimization phase to apply next, given characteristics of the current,
partially optimized code. They evaluated their method in a dynamic compilation setting, using Java. 
Singh et al. \cite{SINGH} used a method similar to ours, where they trained an artificial neural network
performance model. However, they focus on large-scale parallel platforms such as the BlueGene/L,
and do not use their model as part of a auto-tuner.
Yigitbasi et al. \cite{YIGITBASI} also adopt a similar approach to us, by building a machine learning based
performance model for MapReduce with Hadoop, and using it in an auto-tuner.
In contrast to these works, our method uses values of tuning parameters to directly predict 
execution time, as part of an auto-tuner, using OpenCL in a heterogeneous setting.

Machine learning approaches have also been used for auto-tuning applications in a heterogeneous setting.
A number of works deals with developing methods to determine whether to execute a kernel on the GPU or
CPU\cite{OGILIVIE,GREWE} and to balance load between the devices\cite{GRASSO,LI_MA}.
Magni et al. \cite{MAGNI} used an ANN model to determine the correct thread coarsening factor, that 
is, the amount of work per thread, based on static code features, for OpenCL on different platforms. In 
another study, they use a nearest neighbor approach to determine the best way to parallelize 
sequential loops for OpenACC\cite{MAGNI2}.
A method to determine whether local memory should be used as an optimization for OpenCL is proposed 
in \cite{HAN}. A random forest based model is trained using millions of synthetic benchmarks, and 
based on manually extracted features of the memory access pattern predicts the speedup if local 
memory is used.
Liu et al. \cite{LIU} focus on how the properties of the program inputs affect the performance of CUDA
programs, and develop a method where a machine learning based algorithm can be used to determine the best
optimization parameters based on the input. In contrast to these works, we develop a performance model that
predicts the execution time based on multiple different tuning parameters, and use it in a auto-tuner.

The two works most closely related to ours are \cite{BERGSTRA,JIA}. In \cite{BERGSTRA}, a model based on
boosted regression trees were used to build an auto-tuner, evaluated with a single GPU benchmark,
filterbank correlation.
The Starchart \cite{JIA} system builds a regression tree model which can be used to partition
the design space, discover its structure and find optimal parameter values within the different regions.
It is then used to develop an auto-tuner for several GPU benchmarks.
In contrast, our work uses a different machine learning model, has more parameters for each kernel, and uses
OpenCL to tune applications for both CPUs and GPUs.

Work has also been done on OpenCL performance portability. Zhang et al. \cite{ZHANG} identify a 
number of parameters, or tuning knobs, which affects the performance of OpenCL codes on different 
platforms, and shows how setting the appropriate values can improve performance. Faberio et al. 
\cite{FABERIO} use iterative optimization to adapt OpenCL kernels to different hardware by picking 
the optimal tiling sizes. Pennycook et al.\cite{PENNYCOOK} take a different approach, and attempt 
to determine application settings which will achieve good performance on different devices, rather 
than optimal performance on any single device.

\section{Background}
\label{background}
One of the currently most popular heterogeneous platforms is the combination of a CPU and a GPU. 
While GPUs traditionally were developed for rendering graphics, they have evolved into general 
purpose programmable, highly parallel accelerators. Here we will only present a brief overview, for 
details on GPU architecture and applications, the reader is referred to \cite{OWENS, BRODTKORB}, and \cite{SMISTAD}, which
include a discussion of OpenCL on AMD and Nvidia GPUs.

GPUs consists of a number of compute units, each of which consists of several processing 
elements\footnote{Here we are adopting the terminology of OpenCL. On Nvidia GPUs, these are known 
as streaming multiprocessors and CUDA cores, on AMD GPUs as compute units and stream processors.}. 
The processing elements of a compute unit work in a SIMD fashion, executing instructions in lock 
step. The largest memory space available is global memory, which resides in slow, off-chip 
DRAM (but is separate from the system's main memory). While the global memory is cached on newer 
GPUs, they also have a fast, on-chip, scratch pad memory, which can be used as a user managed cache. 
In addition, they have texture memory, which is optimized for access patterns with 2D and 3D 
spatial locality, and constant memory designed to allow for high performance when accessed by many 
threads concurrently.
 
\subsection{OpenCL}
OpenCL\cite{OPENCL} is a standard for heterogeneous computing, which makes it possible to write 
code once, and execute it on different devices, including CPUs and GPUs. The code is organized into 
host code and kernels. The host code executes as a normal CPU program, and sets up and launches the 
kernels on a device (which might be the same CPU the host code is executing on). Kernels are 
executed in parallel by a number of threads known as work items, which are organized into work 
groups. If executed on a GPU, the work groups are typically mapped to compute units, and the work 
items to processing elements, on the CPU they are mapped to the CPU cores. A number of logical memory spaces 
exists: local memory (mapped to the fast on-chip memory on GPUs), image memory 
(mapped to the GPU texture memory), and constant memory (mapped to the hardware 
constant memory on the GPU). On the CPU, all of these memory spaces are typically mapped to main 
memory.
 
\subsection{Machine Learning}

This section will provide a basic introduction to machine learning with a focus on artificial neural 
networks, more details can be found in e.g. \cite{HASTIE}. Machine learning algorithms can 
be broadly divided into unsupervised and supervised learning. In unsupervised learning, no output 
labels are given for the input, instead, the algorithm attempts to discover structure in the input 
data itself. In supervised learning on the other hand, the algorithm is given example input and 
output pairs, which are used to build a model which can later predict the output for unseen input. 
If the outputs are categories, this is known as classification, if they are real numbers, as 
regression. In both cases it can be viewed as form of function approximation.

\begin{figure}[htbp]
    \centering
    \includegraphics[width=0.4\textwidth]{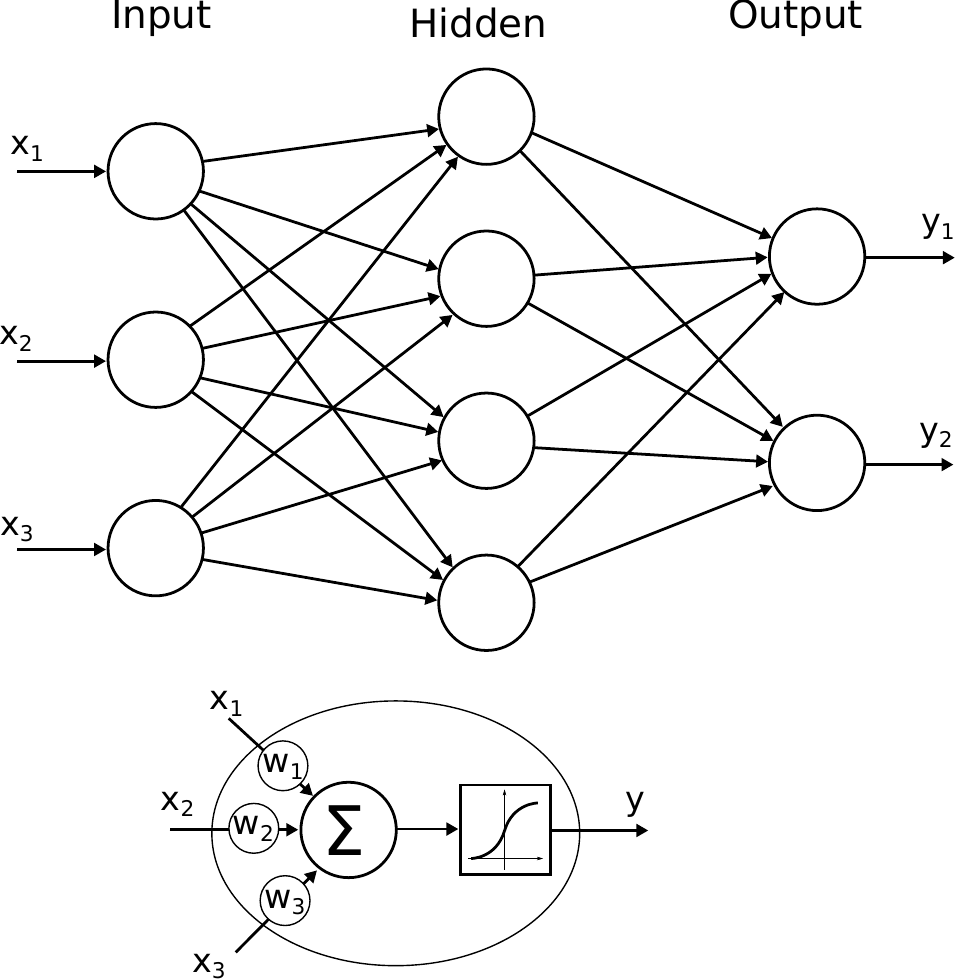}
    \caption{Neural network with 3 input neurons, a single hidden layer 
with 4 neurons, and two output neurons. Bellow, a single neuron with three inputs.}
    \label{nn}
\end{figure}

Artificial neural networks are a supervised machine learning algorithm which can be used for both 
classification and regression. They are built up of artificial neurons, which have multiple inputs 
and a single output. The output of a neuron is found by first computing a weighed sum of the inputs, 
and then passing the result to an activation function, such as the sigmoid or threshold function. A 
neural network can be built by connecting the inputs and outputs of multiple neurons. Some neurons 
have unconnected inputs, these are the input neurons, while others have unconnected outputs, these 
are the output neurons. By presenting values to the input nodes, the network can compute the output 
by letting the values propagate through the network. A complete network, and a detailed view a 
single neuron, is shown in Figure~\ref{nn}.

The weights of the neurons determine what the network computes and must be fitted to the example 
data, a process known as training. Several training algorithms exists, they all initialize the 
weights to random values, and attempt to adjust them so that the values computed by the network for 
the example inputs matches the example outputs.

The topology and activation function used in a network must be adjusted manually to the problem at 
hand. These factors can greatly affect the performance of the network, but little knowledge about 
how to pick values exists, and experimentation is often required.

\section{Machine Learning based auto-tuning}
\label{work}
\begin{figure}[htbp]
    \centering
    \includegraphics[width=0.45\textwidth]{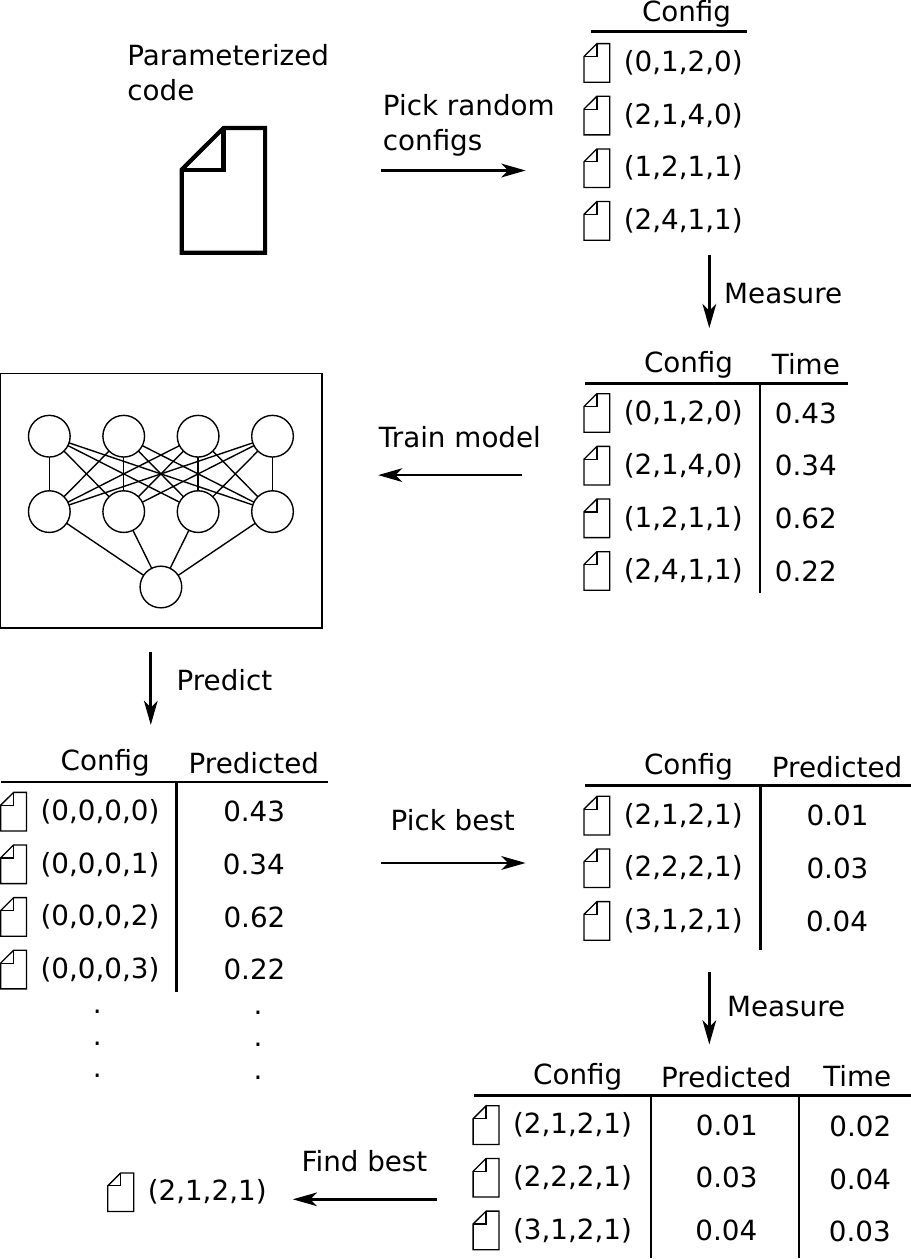}
    \caption{Overview of our auto-tuning approach. Execution times for a random set of
        configurations are measured, and used to build a model, which is then
        used to pick interesting configurations for exhaustive search.
        }
    \label{overview}
\end{figure}

Figure~\ref{overview} illustrates our auto-tuning method. We start with parameterized benchmarks. 
The parameters form a space of possible implementations, and from this space we pick samples which 
are used to build the machine learning based model. The model is then used to predict the execution 
time for all the possible configurations. In a second stage, the $M$ configurations with the lowest 
predicted execution times are found, and their actual execution times are measured. The best of 
these configurations is then found, and returned by the auto-tuner. If the model is sufficiently 
accurate, the optimal configuration will be among those found in the second stage, and therefore 
returned by the auto-tuner.

In the following we will cover these steps in more detail.

\subsection{Code parameterization and Candidate Generation}

We used three benchmarks for our experiments, \texttt{convolution}, \texttt{raycasting} and 
\texttt{stereo}, described in Table~\ref{benchmarks}. The code of the benchmarks 
were parameterized with tuning parameters, to make it possible to generate multiple candidate 
implementations, with one candidate for each tuning parameter configuration. These candidates are 
all functionally equivalent, but the different values of the tuning parameters causes their 
performance to vary.

The tuning parameters determine whether or not various optimizations are applied, as well as the 
value of performance critical variables. Examples include whether or not to manually cache values 
in local memory, how much work should be assigned to each thread, and whether or not to perform 
loop unrolling. An overview and description of all the tuning parameters used, and their possible 
values, can be found in Table~\ref{param_table}.

The optimal values for the parameters depends on the device being used, and possibly the input to the algorithm.
Furthermore, since the parameters are not independent, the best values cannot be found by varying the values of one 
parameter at a time. As can be seen from Table 1, the size of the parameter spaces for are 131K, 
655K and 2359K for \texttt{convolution}, \texttt{raycasting} and \texttt{stereo} respectively.

The parameters are implemented either using preprocessor macros in the kernel code, or using 
variables set on the host prior to kernel launch. The loop unrolling in \texttt{convolution} and
\texttt{stereo} is implemented using OpenCL driver pragmas, while in \texttt{raycasting}, it is done
manually, using preprocessor macros. Several of the parameters deal with storing data structures in
different memory spaces. These parameters can in general be combined in any way, for instance, if
both image and local memory is used, the data structure will first be stored in image memory, and
then be manually cached in local memory.

\begin{table}
\caption{Benchmarks used}
\begin{tabularx}{0.45\textwidth}{|l|X|}
\hline
\textbf{Benchmark} & \textbf{Description} \\ \hline
\texttt{convolution} & convolution of 2048x2048 2D image with 5x5 box filter, example of stencil computation. \\ \hline
\texttt{raycasting} & Volume visualization generating 1024x1024 2D image from 512x512x512 3D volume data. \\ \hline
\texttt{stereo} & Computing disparity between two 1024x1024 stereo images to determine distances to objects.  \\ \hline
\end{tabularx}
\label{benchmarks}
\end{table}

\begin{table}[h]
\caption{Parameters used for the benchmarks and their possible values. The parameters 
listed under \textbf{all} are present in all the benchmarks.}
{\footnotesize
\begin{tabular}{|l|l|}
\hline
\multicolumn{2}{|c|}{\textbf{all}} \\ \hline
\textbf{Parameter} & \textbf{Possible values} \\ \hline
Work-group size in x dimension & 1,2,4,8,16,32,64,128 \\ \hline
Work-group size in y dimension & 1,2,4,8,16,32,64,128 \\ \hline
Output pixels per thread in x dimension & 1,2,4,8,16,32,64,128 \\ \hline
Output pixels per thread in y dimension & 1,2,4,8,16,32,64,128 \\ \hline
\hline
\multicolumn{2}{|c|}{\textbf{convolution}} \\ \hline
Parameter & Range \\ \hline
Use image memory &  0,1 \\ \hline
Use local memory &   0,1 \\ \hline
Add padding to image  &  0,1 \\ \hline
Interleaved memory reads &  0,1 \\ \hline
Unroll loops  & 0,1 \\ \hline
  \hline
\multicolumn{2}{|c|}{\textbf{raycasting}} \\ \hline
\textbf{Parameter} & \textbf{Possible values} \\ \hline
Use image memory for data & 0,1 \\ \hline
Use image memory for transfer function & 0,1 \\ \hline
Use local memory for transfer function  & 0,1 \\ \hline
Use constant memory for transfer function & 0,1 \\ \hline
Interleaved memory reads  & 0,1 \\ \hline
Unroll factor for ray traversal loop & 1,2,4,8,16 \\ \hline
 \hline
\multicolumn{2}{|c|}{\textbf{stereo}} \\ \hline
\textbf{Parameter} & \textbf{Possible values} \\ \hline
Use image memory for left image & 0,1 \\ \hline
Use image memory for right image & 0,1 \\ \hline
Use local memory for left image  & 0,1 \\ \hline
Use local memory for right image & 0,1 \\ \hline
Unroll factor for disparity loop  & 1,2,4,8 \\ \hline
Unroll factor for difference loop in x direction & 1,2,4 \\ \hline
Unroll factor for difference loop in x direction  & 1,2,4 \\ \hline
\end{tabular}
}
\label{param_table}
\end{table}

\subsection{Model building}
The model we build should be able to predict the execution time of a benchmark given the tuning 
parameter configuration. To do this, we run the benchmarks on several randomly chosen 
parameter configurations and record the execution time. These input-output pairs, or 
training samples, are then used to build, or train, a model. Using machine learning terminology, 
this is known as supervised learning.

Multiple supervised learning algorithms exist. We have used artificial neural networks (ANN) due to 
their good predictive power, ability to handle arbitrary functions, and ability to handle noisy input robustly.
A significant drawback of ANNs is, however, the opaqueness of the resulting model, which 
makes it difficult to interpret, and hard to gain deeper insights into how the different parameters 
interact, and contribute to the final performance.

Through experimentation, we found that a network with a single hidden layer with 30 neurons using 
sigmoid activation functions gave good performance.

Additionally, we used a technique know as \emph{bagging}\cite{ZHOU} to further increase the performance of the
model. Rather than using all the training data to build a single neural network, we split it into
$k$ parts and build $k$ networks, each trained using all the data except one of the parts. During
prediction, we feed the input to all the networks, and then take the mean of their outputs as the
final output. We found that this increased the accuracy of the predictions. We have used a value of 11 for $k$.

During ANN training, the weights are adjusted to minimize the mean
squared error between the predictions and the actual output. In our case, this causes problems since we use the
ANN to predict the execution time directly, and are therefore interested in minimizing
the relative, rather than the absolute error. To resolve this problem, we take the logarithm of the
execution times before training the neural network. The neural network then predicts the logarithm
of the execution time, and attempts to minimize the mean squared error when comparing with the
logarithm of the actual execution time. This works since reducing the absolute error of the logarithm of two values is equivalent to reducing the
relative error of the values directly.

A challenge specific for this kind of data is invalid parameter configurations, that is, 
configurations for which the corresponding code cannot be run. This is typically because the 
resulting code uses too many resources, for instance, some devices places restrictions on how large 
work-groups can be, or how much local memory is available. If the specific device is known, most of 
the invalid configurations can be determined statically, but in some cases it is necessary  to 
attempt to compile and run the kernels. We deal with this issue by simply ignoring these 
configurations when training the model. 

\subsection{Prediction and Evaluation}
After the model is built, the optimal parameter configuration may be found by simply 
predicting the execution time for all possible configurations, and picking the best one. This 
remains feasible despite large parameter spaces since it is orders of magnitude faster to evaluate 
the model than to execute the actual benchmarks.

However, since the model is not perfectly accurate, it is unlikely that the configuration with the 
lowest predicted execution time is the one with the lowest actual execution time. Our auto-tuner
therefore includes a second stage where the model to find interesting subspaces of the 
parameter space which are small enough to be searched exhaustively, and are likely to contain the 
actual optimal configurations.

In practice, we do this by picking the $M$ configurations with the lowest predicted execution times. We 
then measure the actual execution times of these configurations, and find the best among them. 
Again, this is not guaranteed to be the globally best configuration since the model may be so 
inaccurate that the globally optimal configuration is not included among the $M$ configurations in 
the second stage. Sometimes these $M$ configurations also include invalid configurations. This is, 
however, not a big problem if the models are trained with enough data.

Our experiments used values for $M$ in the range 10 - 200, with good results. However, by
making assumptions about the distribution of the execution times, as well as the distribution of
prediction errors this ad-hoc method could be replaced with a more principled one where one could
determine values for $M$ so that the samples in the second stage contains the optimal one with a
given probability.

\section{Results}
\label{results}

To evaluate our method, we implemented 3 parameterized benchmarks in OpenCL, \texttt{stereo}, 
\texttt{convolution} and \texttt{raycasting}, see  
Table~\ref{benchmarks}. Descriptions of the parameters for all the benchmarks can be found in
Table~\ref{param_table}. For the experiments, we used an Nvidia K40 GPU, an AMD Radeon HD 7970 GPU and an
Intel i7 3770 CPU.

The time required to train the ANN models is significant, but small compared to the cost of 
gathering the training data. For example, for the \texttt{convolution} benchmark on the Nvidia GPU, 
training the model with 2000 samples takes about 1 minute, gathering the data takes about 30 
minutes. The time required to gather data is so high because it does not only include the time 
for the kernels themselves, but also the overhead of compiling the kernels, as well as time wasted 
attempting to compile and launch kernels with invalid configurations.

\begin{figure}[htpb]
    \centering
    \includegraphics[width=0.45\textwidth]{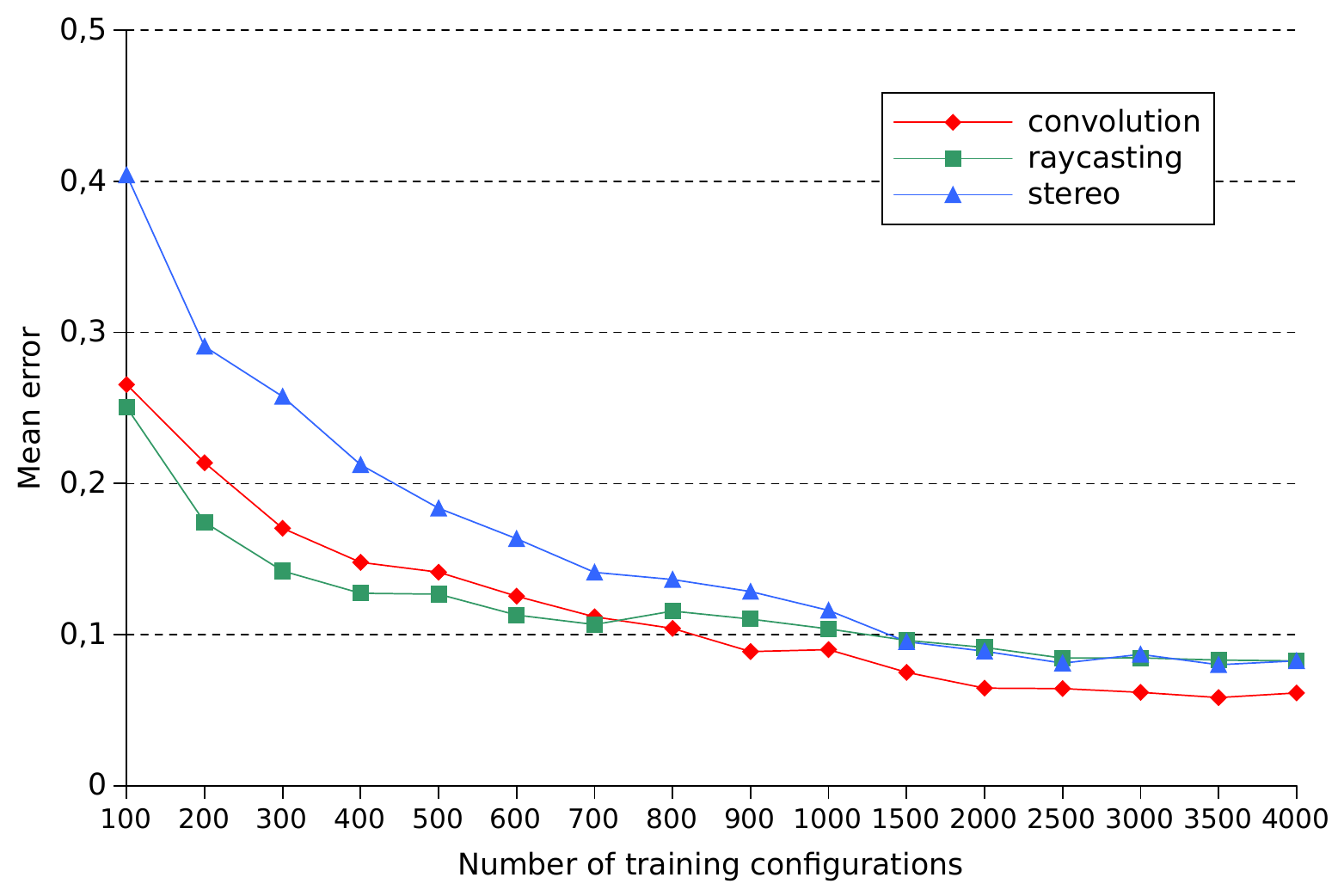}
    \caption{Mean prediction error for the different benchmarks for increasing number of training 
samples, on the Intel i7.}
\label{accuracy_intel}
\end{figure}

\begin{figure}[htpb]
    \centering
    \includegraphics[width=0.45\textwidth]{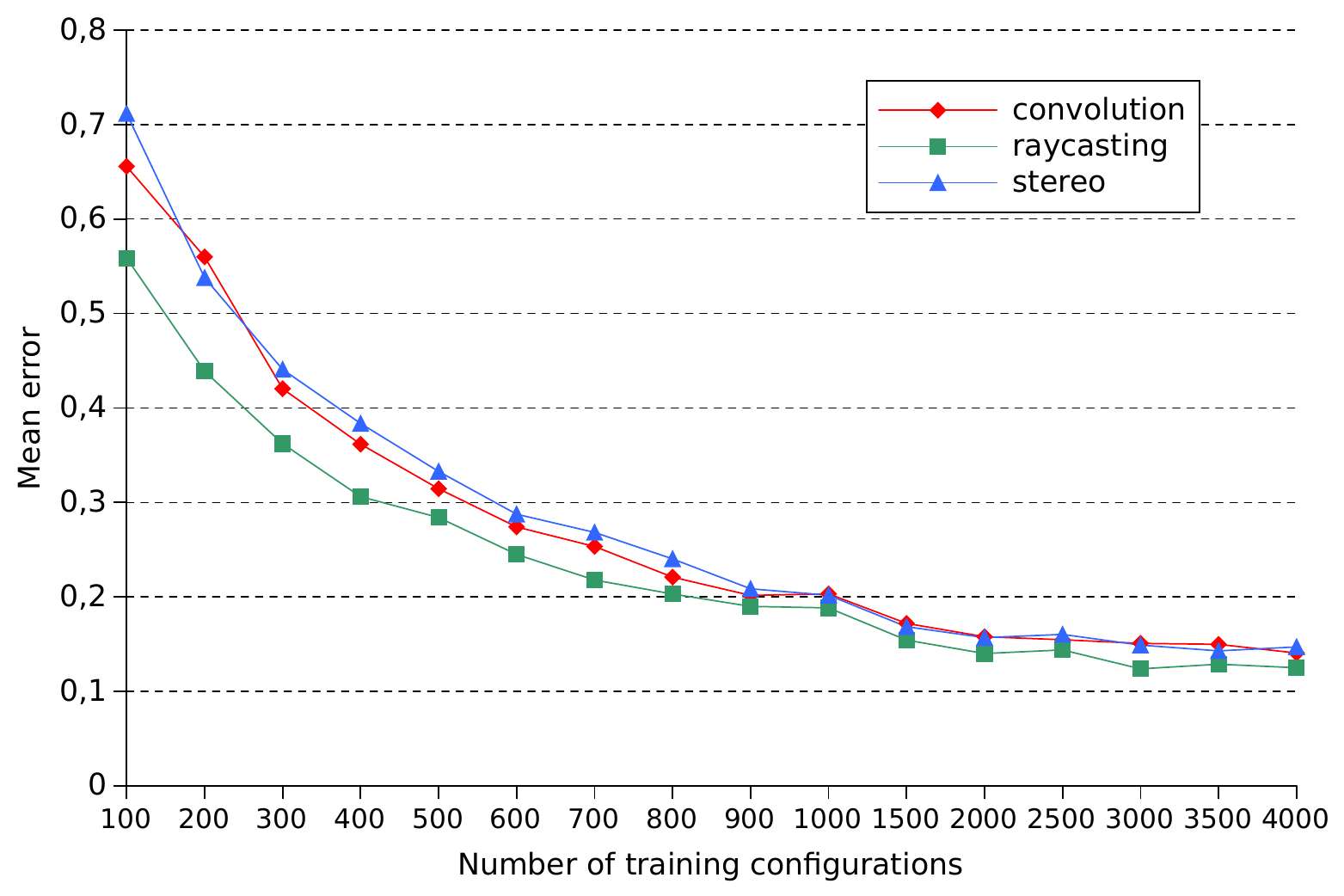}
        \caption{Mean prediction error for the different benchmarks for increasing number of 
training samples, on the Nvidia K40.}
\label{accuracy_nvidia}
\end{figure}

\begin{figure}[htpb]
    \centering
    \includegraphics[width=0.45\textwidth]{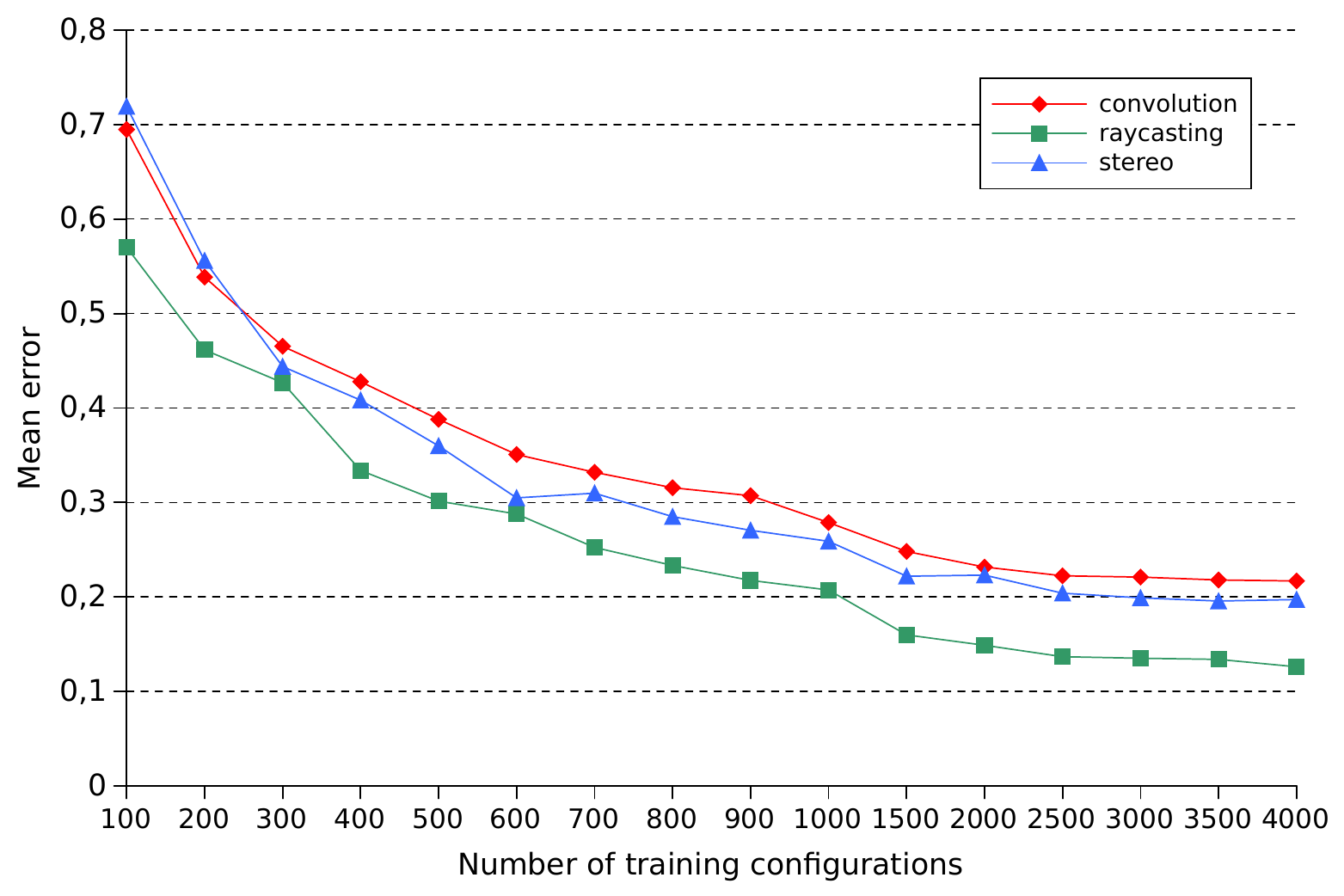}
        \caption{Mean prediction error for the different benchmarks for increasing number of 
training samples, on the AMD 7970.}
\label{accuracy_amd}
\end{figure}

To evaluate the accuracy of the models created, we compared the predictions against actual 
execution times for valid parameter configurations not used during training. This was repeated 
for 
neural networks trained using an increasing number of configurations. Since the output of the model 
depends 
on the particular configurations used during training, as well as the random initial weights of the
neural network, we built several neural networks using different configurations for each training
size and report the mean of the output for all these networks. The results are shown in figures 
\ref{accuracy_intel}, \ref{accuracy_nvidia} and \ref{accuracy_amd} for the Nvidia, Intel and AMD 
devices respectively.

As is shown, the mean relative error decreases as more samples are used to train the models, but 
stabilizes or decreases much more slowly after around 1000-2000 samples, for all devices and benchmarks.
The accuracy on the Intel CPU is noticeably better than for the GPUs, the relative mean accuracy is
6.1\% - 8.3\% for 4000 training configurations on the CPU, the corresponding 
numbers are 12.5\%-14.7\% and 12.6\%-21.2\% for the Nvidia and AMD GPUs respectively. The
performance of the different benchmarks is relatively similar on the Intel CPU and Nvidia GPU, on
the AMD, \texttt{raycasting} performs significantly better than \texttt{convolution} and
\texttt{stereo}.

\begin{figure}[htpb]
    \centering
    \includegraphics[width=0.45\textwidth]{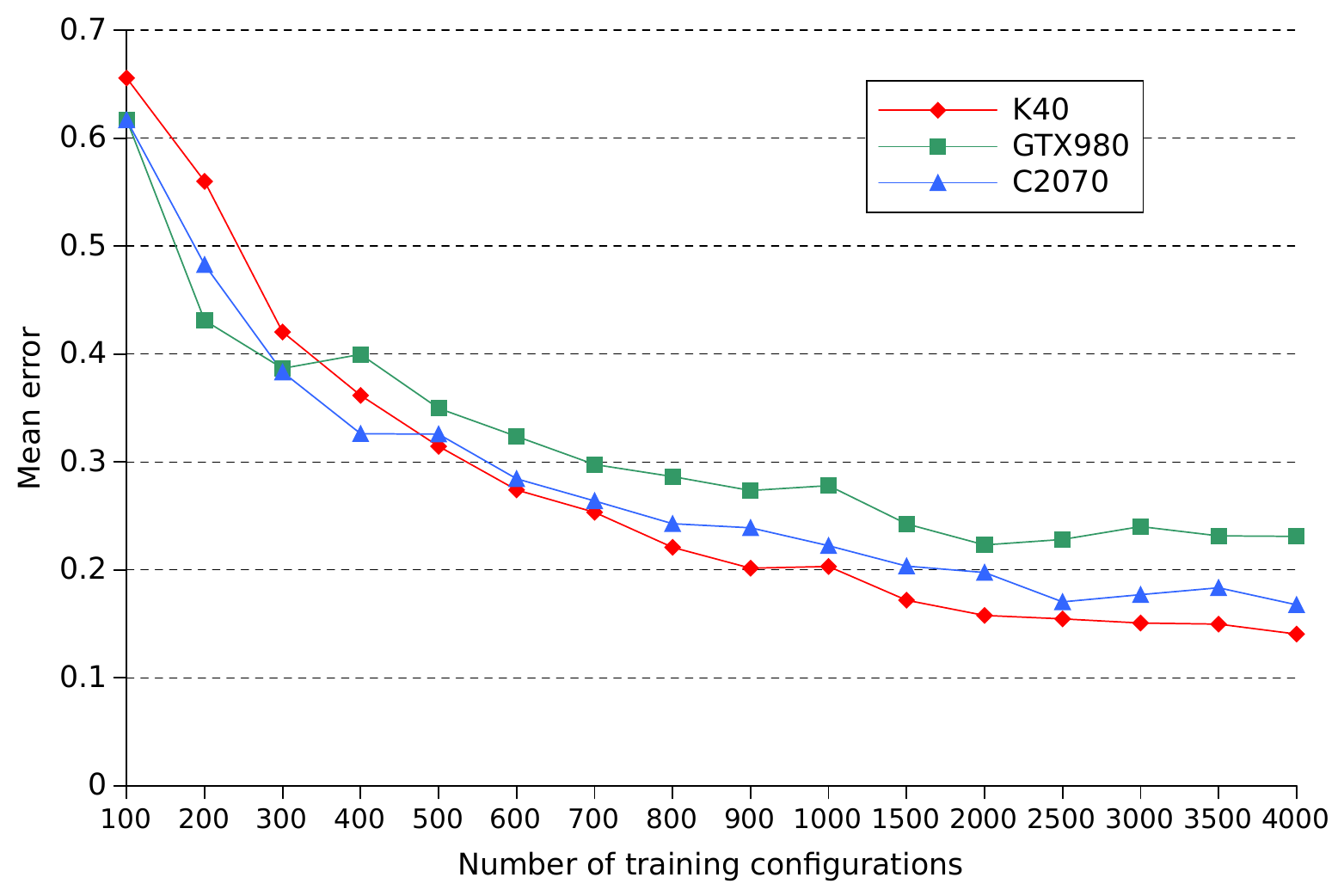}
        \caption{Mean prediction error for convolution on different Nvidia GPUs.}
\label{different_nvidia}
\end{figure}

We also investigated the performance on different devices from the same vendor, 
Figure~\ref{different_nvidia} shows accuracy for the convolution benchmark for three different Nvidia GPUs,
a C2070, a K40 and a GTX980, representing the Fermi, Kepler and Maxwell architectures respectively. As the
figure shows, the accuracy is similar for the K40 and c2070, while slightly worse for the GTX980.

\begin{figure}[htpb]
    \centering
    \includegraphics[width=0.45\textwidth]{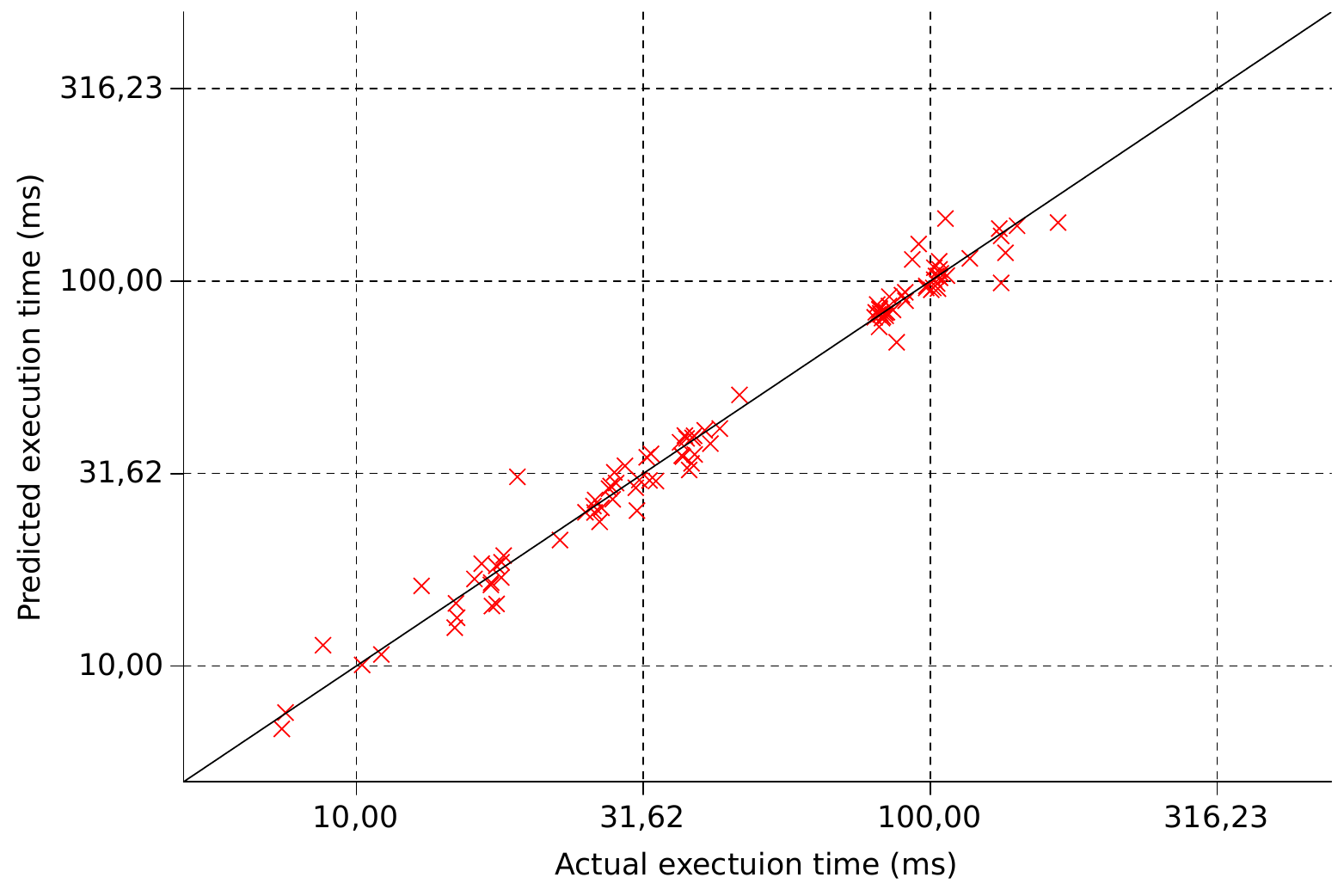}
    \caption{Predicted versus actual execution times, on the Intel i7. Note the logarithmic scales.}
\label{scatter_intel}
\end{figure}

\begin{figure}[htpb]
    \centering
    \includegraphics[width=0.45\textwidth]{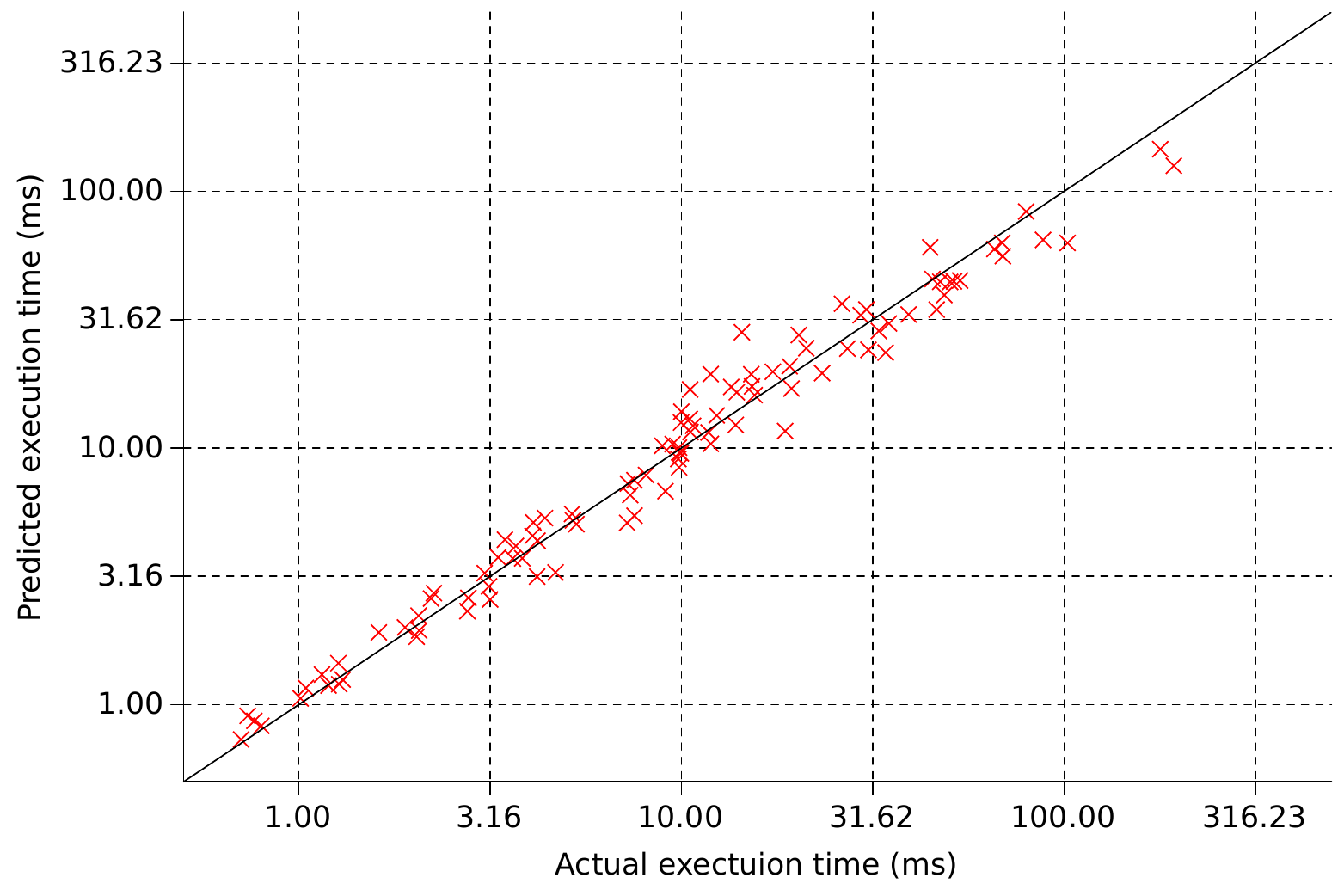}
    \caption{Predicted versus actual execution times, on the Nvidia K40. Note the logarithmic 
scales.}
\label{scatter_nvidia}
\end{figure}

\begin{figure}[htpb]
    \centering
    \includegraphics[width=0.45\textwidth]{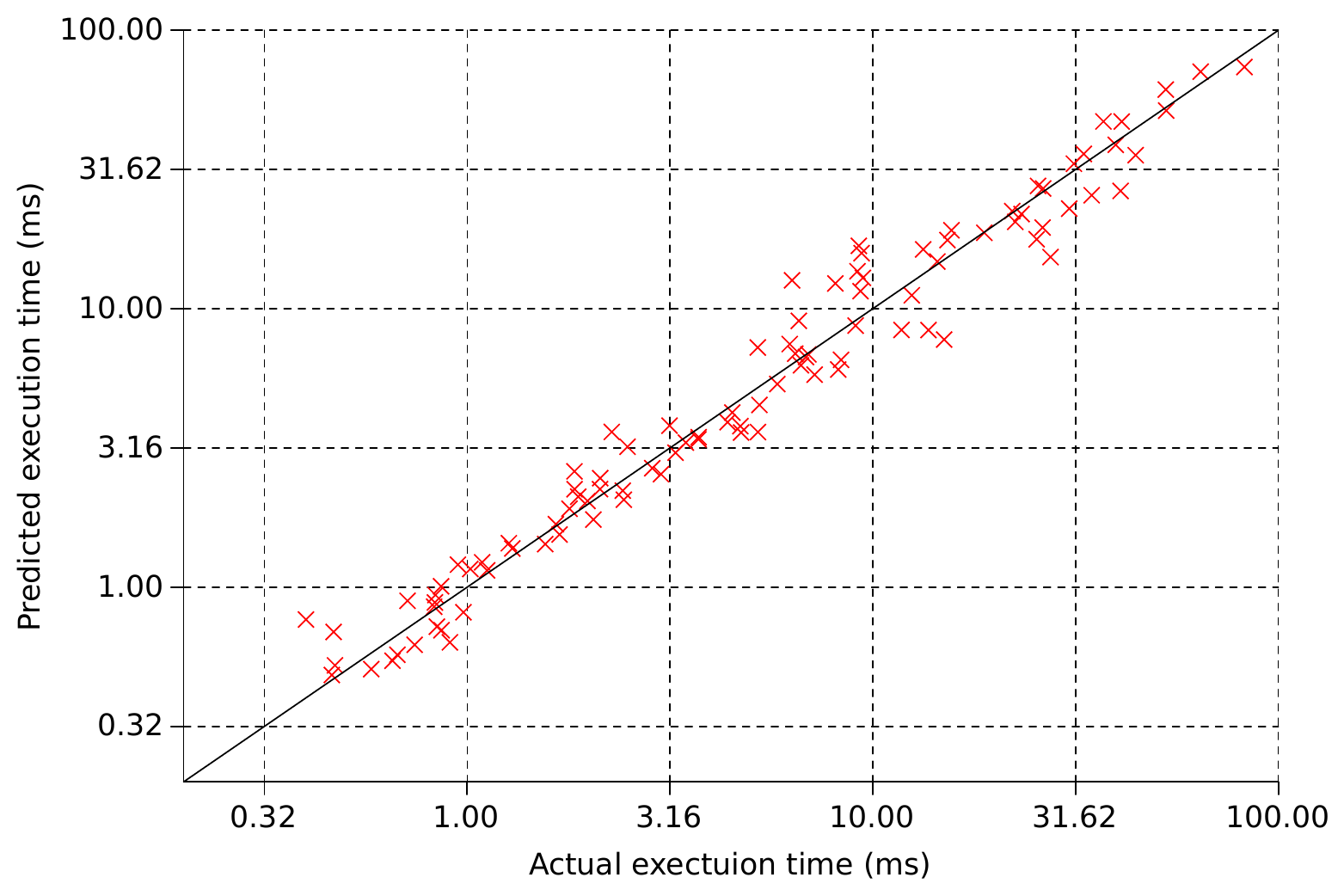}
    \caption{Predicted versus actual execution times, on the AMD 7970. Note the logarithmic scales.}
\label{scatter_amd}
\end{figure}

Another way to visualize the accuracy is to scatter plot the actual versus the predicted
execution times. This is done for the convolution benchmark in figures \ref{scatter_nvidia}, \ref{scatter_intel} and
\ref{scatter_amd}, for the Nvidia, Intel and AMD devices, respectively. The figures show 100
configurations not used during training. In this case, the results are not the average of multiple
models. As the figures show, there is a good match between predictions and reality. The clustering of the
Intel data is caused by the values of the local and image memory parameters. Here, using image memory
without local memory results in a significantly worse performance compared to all other
combinations of these parameter values.

For the \texttt{convolution} benchmark, the parameter space is fairly small compared to the two 
other benchmarks, and it was therefore possible to measure the actual execution times of all possible 
configurations. This allows us to evaluate our auto-tuners ability to find good configurations, since
we can compare against the known globally optimal configuration.

\begin{figure}[htpb]
    \centering
    \includegraphics[width=0.45\textwidth]{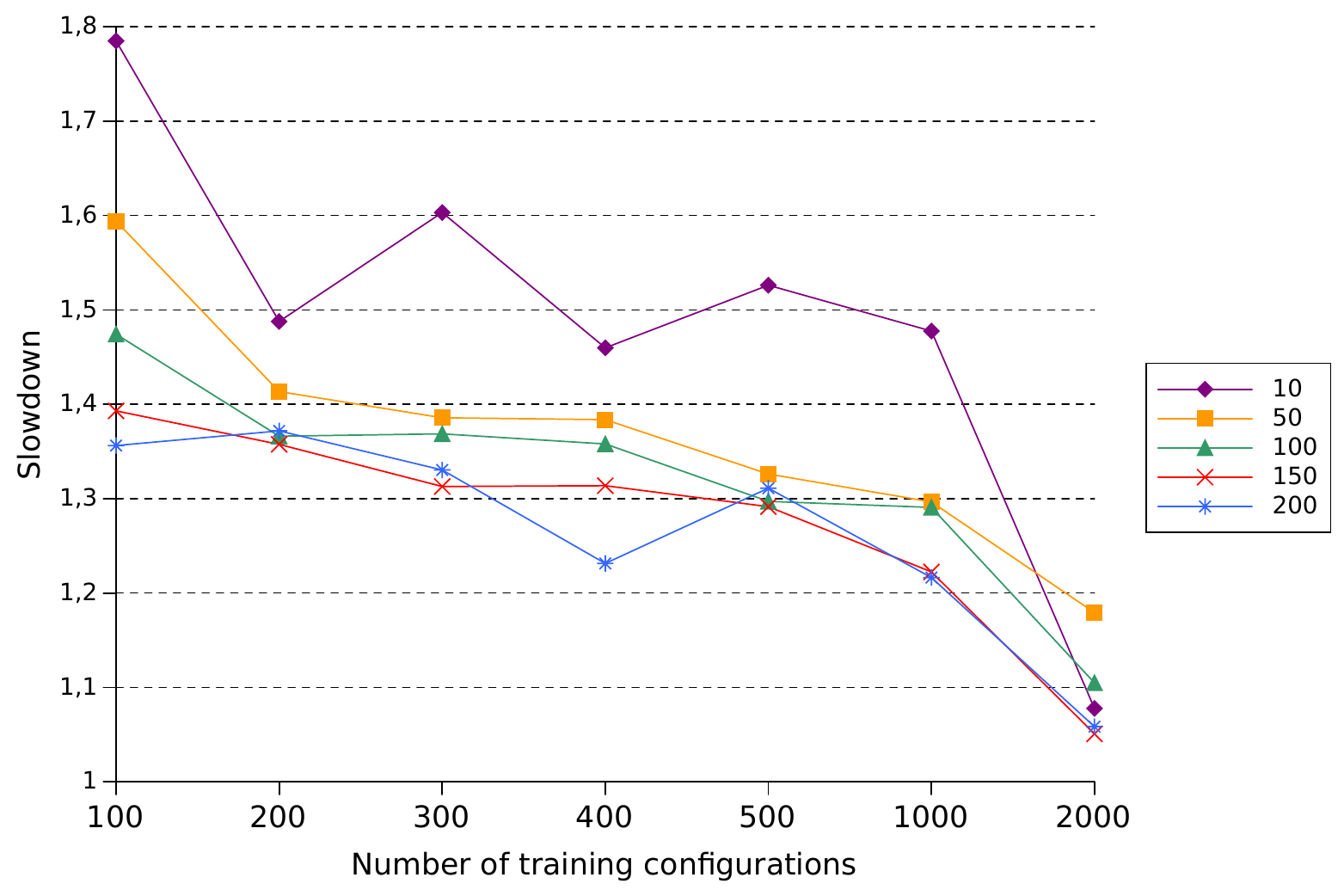}
        \caption{Average slowdown of results from our auto-tuner compared to the globally optimal 
solution, Nvidia K40.}
\label{compare_nvidia}
\end{figure}

\begin{figure}[htpb]
    \centering
    \includegraphics[width=0.45\textwidth]{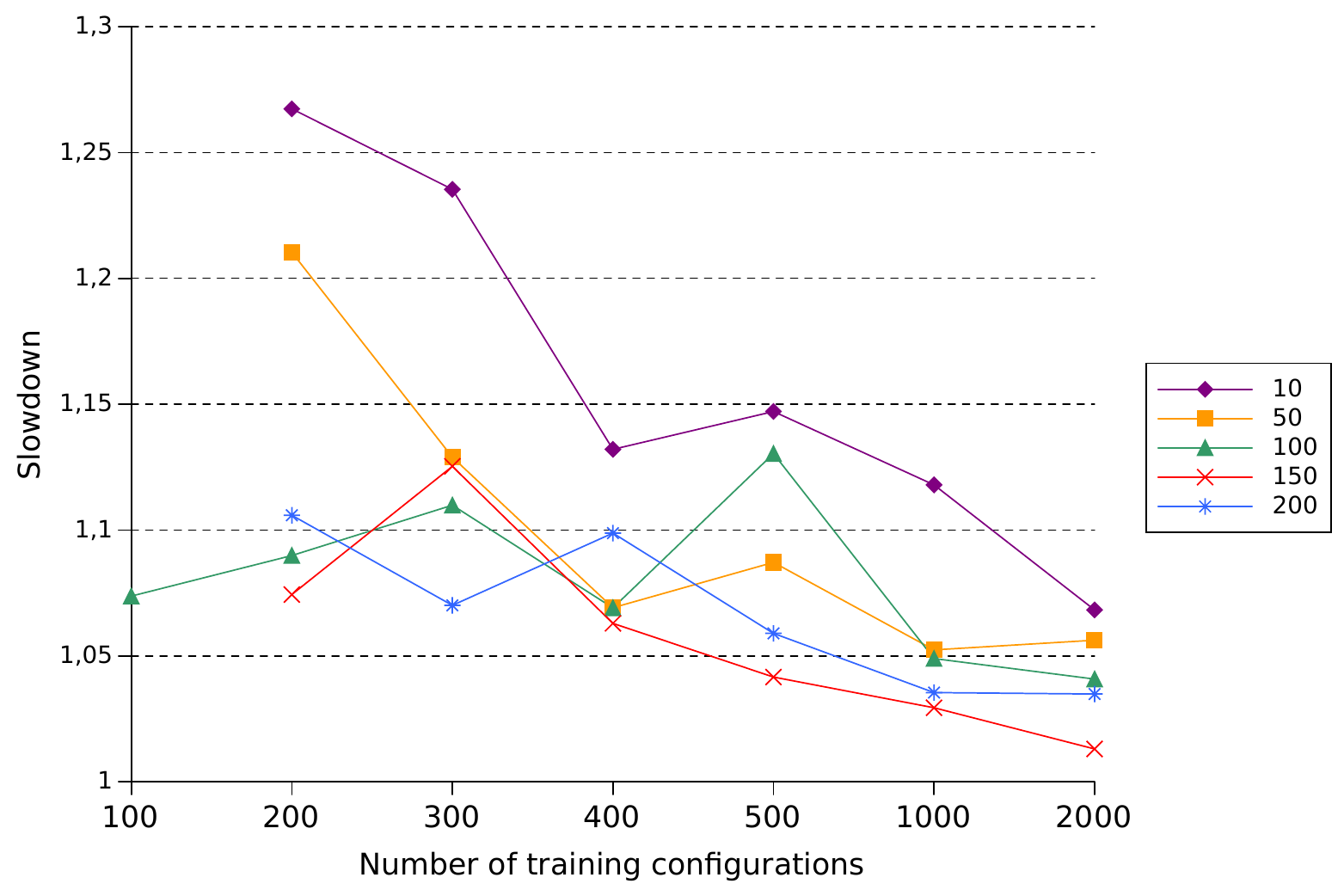}
        \caption{Average slowdown of results from our auto-tuner compared to the globally optimal 
solution, on the
Intel i7. Some results missing due to invalid configurations.}
\label{compare_intel}
\end{figure}

\begin{figure}[htpb]
    \centering
    \includegraphics[width=0.45\textwidth]{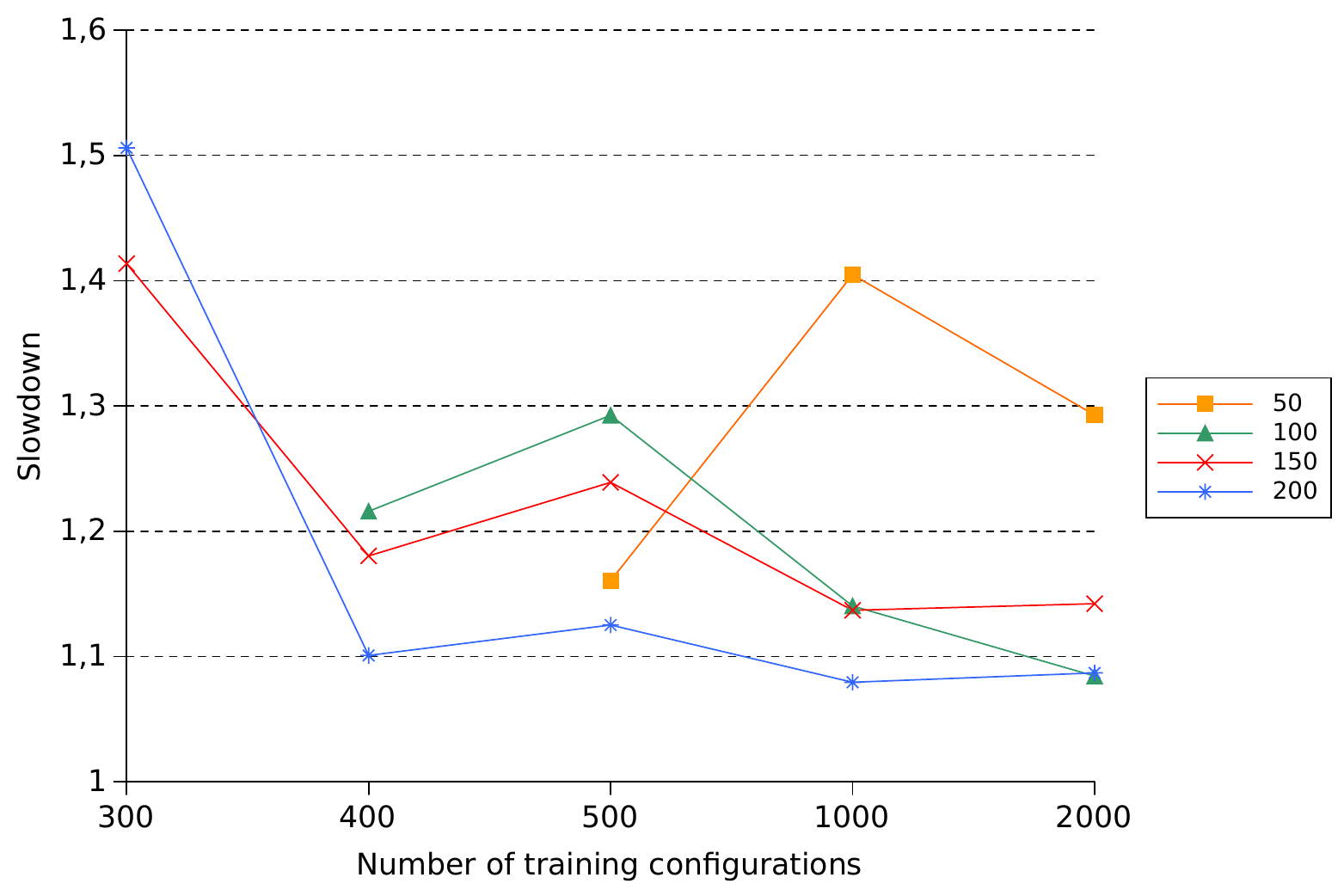}
        \caption{Average slowdown of results from our auto-tuner compared to the globally optimal 
solution, on the
AMD 7970. Some results missing due to invalid configurations.}
\label{compare_ati}
\end{figure}

In figures~\ref{compare_nvidia},~\ref{compare_intel} and \ref{compare_ati} we have varied both the number of configurations
used for training the model, and the number $M$ of configurations used in the second stage of our
auto-tuner.  We expect best results when both of these are high. As above, we built several networks
for each combination, and report the mean of the results. Some results are missing, due to a high
number of invalid configurations being predicted.

As can be seen, out auto-tuner is able to find good configurations. E.g. when we use 2000 configurations
in the first stage, and 200 in the second stage, we are able to find configurations which on average
are 3.5\%, 5.8\% and 8.7\% slower than the global optimum, for the Intel, AMD and Nvidia devices,
respectively, after evaluating only 1.7\% of the possible configurations. In some cases, we are even able
to find the global optimums, but poorer results in other cases pull the averages down. When fewer
configurations are used, the results are worse, when 500 configurations are used in the first stage,
and 100 in the second, we find solutions on average 13.0\%, 29.3\% and 29.7\% slower than the optimum for the Intel, AMD and
Nvidia devices, respectively.

\begin{figure}[htpb]
    \centering
    \includegraphics[width=0.45\textwidth]{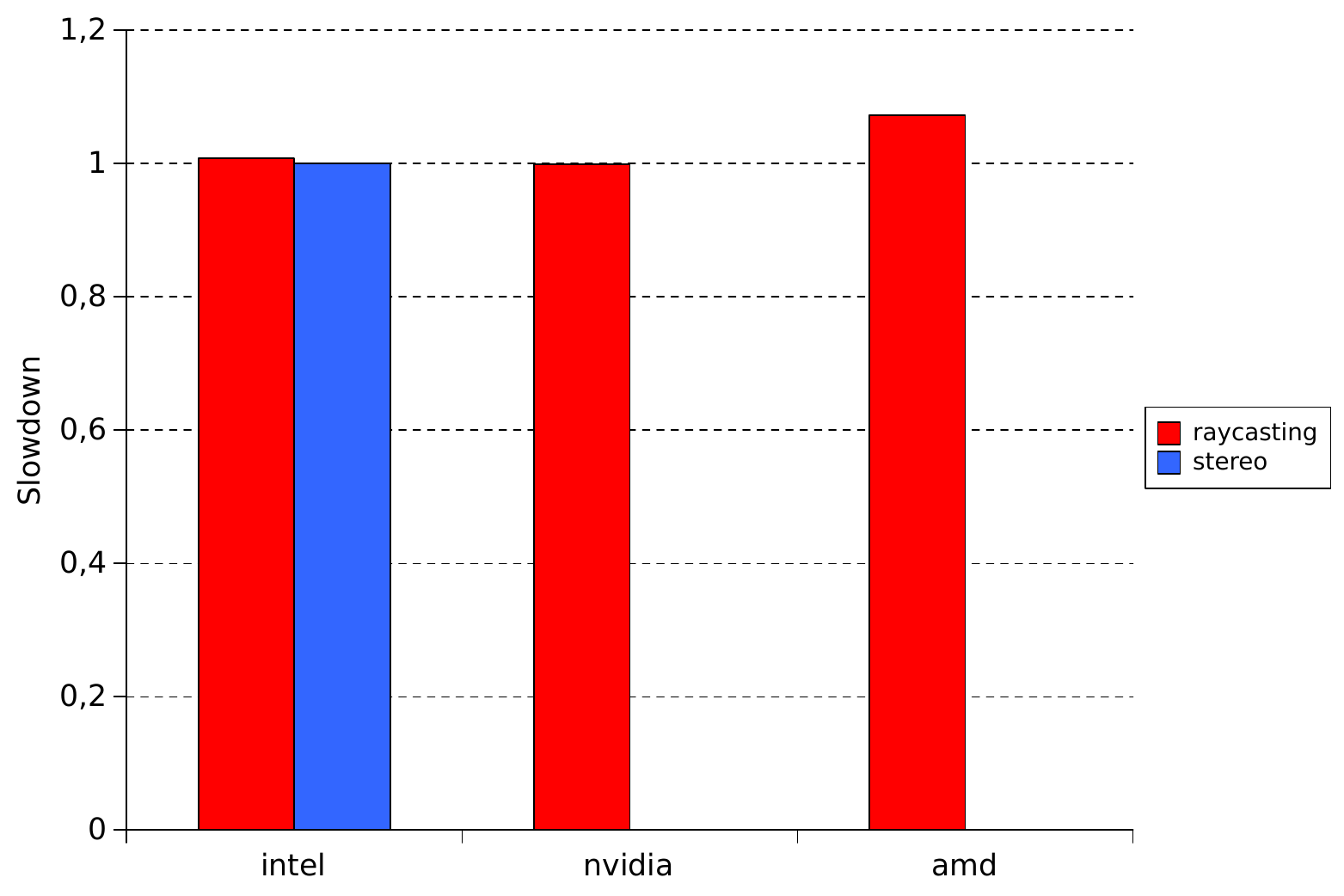}
        \caption{Average slowdown of results from auto-tuner compared to the globally optimal 
solution, for various values of $N$ and $M$, on the AMD 7970. Some results missing due to invalid configurations.}
\label{stereo_ray}
\end{figure}

For the \texttt{stereo} and \texttt{raycasting} benchmarks, the parameter spaces are so large that
time constrains prevented us from exhaustively evaluating all configurations. The best parameter
configurations are therefore not known, making performance evaluation harder. We have, however, measured the execution time of 50K random configurations for
both benchmarks, and compared the best output found to the output of our auto-tuner. 
The results are shown in Figure~\ref{stereo_ray}, here we have used 3000 configurations in the first
stage, and 300 configurations in the second stage. This corresponds to  0.5\% and 0.1\% of the
configurations spaces for \texttt{raycasting} and \texttt{stereo}, respectively. As the figure
shows, we are able to find good configurations, in some cases slightly better than the best among
the 50K random samples. We would like to reemphasize that these are only preliminary
results, since we don't know the truly best configuration. Since the model predicted mostly
invalid configurations for the stereo benchmark on the GPUs, we do not report any
results for these cases.

\section{Further Discussions}
\label{discussion}
First there is the issue of the varying accuracy achieved for the different devices. In particular,
the mean relative error for the Intel CPU is between 6.1\% and 8.3\%, while the corresponding 
numbers are 12.5\%-14.7\% and 12.6\%-21.2\% for the Nvidia and AMD GPUs respectively. One possible explanation is that the memory
related parameters may have less effect on the CPU than the GPUs, since all the logical memory spaces
are mapped to the same physical memory on the CPU, as described in Section~\ref{background}. An exception here is the effect which causes the
clustering of the \texttt{convolution} data, as described in Section~\ref{results}. This effect is not present
in the other benchmarks, and may be the reason why \texttt{convolution} has best accuracy.Furthermore, there are
fewer invalid configurations on the CPU, which increases the accuracy. Finally, while the problem
sizes has been adjusted to partly compensate for this, the execution times on the CPU are generally
longer, potentially making the timing measurements more reliable.

Secondly, there is the issue of the varying performance for the different benchmarks on the AMD GPU.
On both the Intel and Nvidia devices, the performance for the different benchmarks are fairly equal.
However, on the AMD GPU, \texttt{raycasting} performs significantly better than \texttt{stereo} and
\texttt{convolution}. This may be related to how the AMD OpenCL driver performs loop unrolling. As
described in Section~\ref{work}, the loop unrolling in \texttt{raycasting} is
done manually, with macros, while the unrolling in the two other benchmarks
relies on the OpenCL driver, which may be more unreliable.

Finally the
current method of simply ignoring invalid configurations during model training
have as a consequence that the model have poor accuracy in the invalid
parameter configuration subspaces. This can cause the model to predict that invalid configurations have
low execution times. In some cases, all the configurations in the second stage
can be invalid, the net effect of which is that the auto-tuner gives no
prediction at all. This can be seen in several of out results, and a better
scheme to deal with this should be developed to improve performance.

\section{Conclusion and Future Work}
\label{conclusion}
We have developed and validated a machine learning based auto-tuning framework for OpenCL. The framework measures 
the performance of several candidate implementations from a parameter configuration space and uses 
this result to build a artificial neural network, which works as a performance model. This model is 
then used to find interesting parts of the configuration space, which are explored exhaustively to 
find good candidate implementations. Our neural network model achieves a mean relative error as low 
as 6.1\% for three different benchmarks executed on three different devices, a 
Intel i7 3770 CPU, an Nvidia K40 GPU and a AMD Radeon HD 7970. The autotuner is able to find good 
configurations, at best only 1.3\% slower than the best configuration.

Future work includes enhancing the performance of the model, in particular with regard to invalid 
configurations, evaluating the model on novel hardware architectures, beyond just CPUs and GPUs, 
and integrating problem parameters into the performance model. Incorporating advanced new features specific to a given architecture\cite{FALCH} will remain challenging.
However, studying multi-GPU systems\cite{SPAMPINATO} and looking into multi-variate analysis\cite{MARTENS} may also be interesting avenues of inquiry.

\section{Acknowledgements}
The authors would like to thank Nvidia's CUDA Research Center program, and NTNU for hardware 
donations, and Malik Khan for helpful discussions.

\bibliographystyle{ieeetr}
\bibliography{referanser}

\end{document}